\documentclass[12pt,preprint]{aastex}

\shorttitle{Supergranulation Waves}
\shortauthors{Schou}

\begin{document}


\title{Solar Supergranulation Waves Detected in Surface Doppler Shift}


\author{J. Schou}
\affil{W. W. Hansen Experimental Physics Laboratory, Stanford University, Stanford, CA 94305-4085}



\begin{abstract}
Recently \citet{giz02} suggested that supergranulation
has a wave-like component. In this paper I show that the same
phenomenon can be observed using surface Doppler shift data,
thereby confirming their observations.
I am also able to measure the dispersion
relation to lower wavenumbers and to extend the results for rotation
and meridional flows beyond $\pm 70^\circ$ latitude.
\end{abstract}




\section{Introduction}
Solar supergranulation, which was first described more than 40 years ago
by \citet{lei}, has remained difficult to explain.
While it appears to be similar to the granulation,
several observations,
such as the anomalous flows obtained using correlation tracking
or similar methods (e.g. \citet{beck00}, in the following BS),
challenge the interpretation as a simple convective phenomenon.
An alternative explanation is that the supergranulation has a wave-like
component and observational evidence for this was recently
reported by \citet{giz02} (GDS in the following).

It has been the cause of some concern
that the wave like character has not been observed
previously, given the many measurements of the rotation rate of the
supergranulation and attempts to determine why it appears to be anomalous.
A likely explanation is that it was obscured by the
limitations of the surface Doppler shift data.

Supergranular motions are predominantly
horizontal near the photosphere.
Let $V_\phi$ be the velocity in the longitudinal ($\phi$) direction
and $V_\theta$ the velocity in the latitude ($\theta$) direction.
The observed Doppler velocity $V$ is then given by
\begin{equation}
V=V_\phi\sin\phi + V_\theta\cos\phi\sin\theta
\end{equation}
under the assumption that the observations are made in the equatorial plane
(i.e. that B0 = 0) at a large distance from the Sun.
For the $V_\phi$ component, in particular, the
projection leads to a large modulation of the signal
as the Sun rotates and to a significant spreading of
the power in the frequency domain.

\section{Data Analysis and Results}

The data used here are the same as those used
in BS, which provides further details.
They consist of a 60 day sequence of
1024$^2$ full disk Doppler velocity images
obtained by the MDI instrument \citep{scherrer} on the SOHO spacecraft,
covering the time period 1996 May 24 to 1996 July 22.
These images were derotated and averaged to a 60 minute cadence,
remapped to a uniform grid in longitude and latitude and binned
down to have a resolution of 1200 points covering 360$^\circ$ of longitude
and 600 points in latitude between $\pm 90^\circ$.
A temporal average was removed as part of this process.

Strips in longitude were extracted,
multiplied by a weighting function,
tracked at a latitude dependent rate, zero padded to 360$^\circ$ by 12$^\circ$
and passed through a 3 dimensional Fourier transform.
The weighting function is a product of a function
of $\theta$ and a function of $\phi$. The function of $\theta$
is 1 for $|\delta\theta| \le 4.5^\circ$, 0 for $|\delta\theta| \ge 5.5^\circ$
and falls off as a cosine bell for
$4.5^\circ \le |\delta\theta|\le 5.5^\circ$,
where $\delta\theta$ is the distance from the target latitude.
Two weighting functions in $\phi$ were applied.
One is optimized for recovering $V_\phi$, the other for recovering $V_\theta$.
The first weighting function is given by
$W_\phi = A_\phi {\rm sign}\phi/ \sqrt{\sin^2 \phi + 0.01}$,
where $A_\phi$ is
an apodization function which is 0 for $|\phi| = 0^\circ$ and for
$|\phi| >80^\circ$, 1 for $ 10^\circ < |\phi| <70^\circ$ and
cosine bell apodized for $0^\circ < |\phi| < 10^\circ$ and
$70^\circ < |\phi| < 80^\circ$.
The second weighting function is given by
$W_\theta = A_\theta / \sqrt{\cos^2 \phi + 0.01}$,
where $A_\theta$ is only apodized between $70^\circ$ and $80^\circ$.
The small bias of 0.01 was added to avoid singularities.

\begin{figure}
\plotone{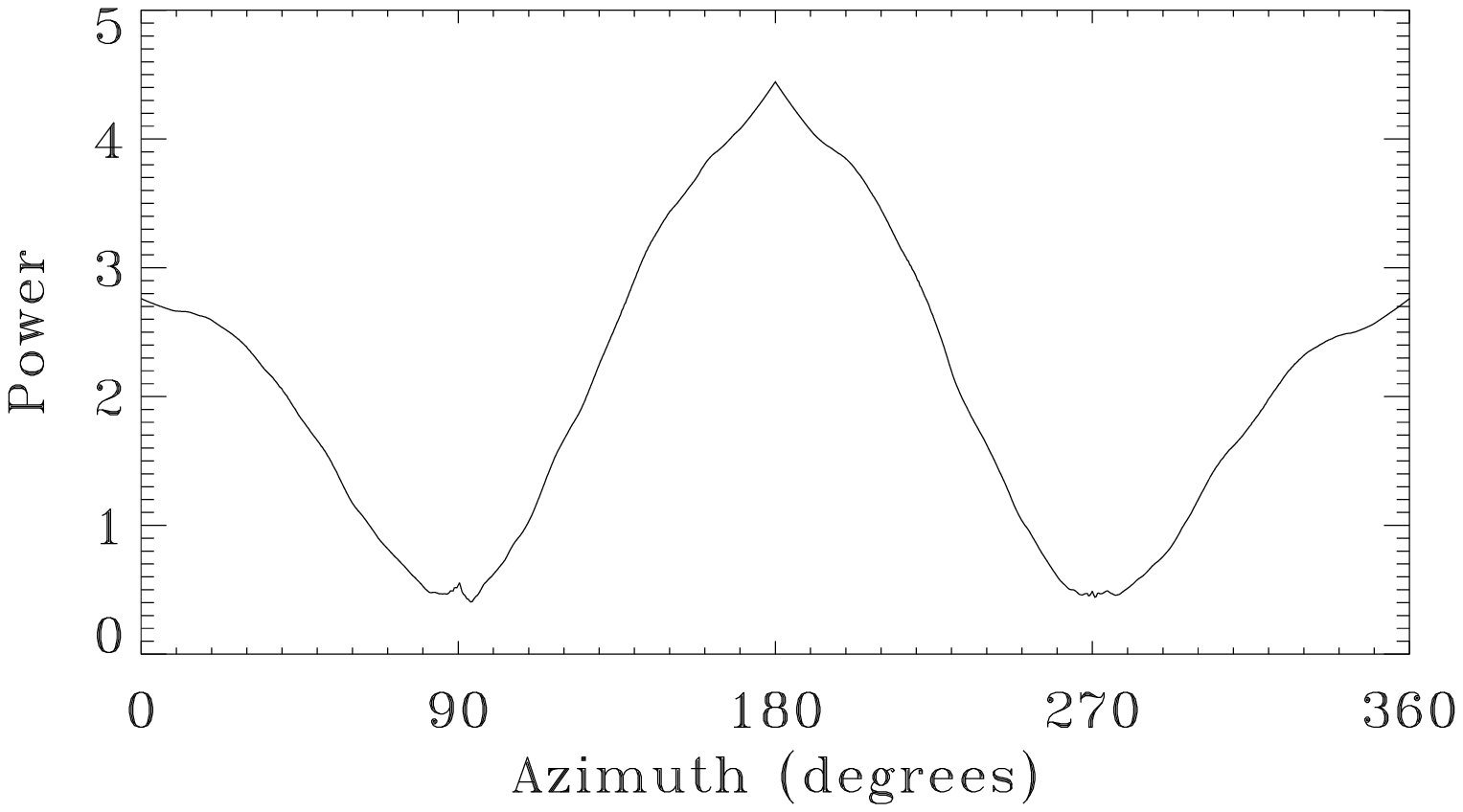}
\caption{
Power (averaged over $30 \le l \le 200$ and $0 < \nu \le 10\mu$Hz)
at the equator as a function of azimuth using the $W_\theta$ weighting.
Prograde is at $180^\circ$ while northward is at $270^\circ$.
\label{pphi}}
\end{figure}

Figure \ref{pphi} shows the power as a function of direction of propagation.
Note that there is significantly
more power in the prograde than in the retrograde direction,
consistent with the findings of GDS.
Also there is little power in the N-S direction where
GDS did observe significant power.
Since the method used by GDS is sensitive to both components of
the velocity this indicates
that the displacement, at least at the surface,
is almost entirely in the direction of propagation.

Given the above
waves in the $\phi$ direction should
show modulation given by $A_\phi$ when using the $W_\phi$ weighting
and waves in the $\theta$ direction should be modulated by
$A_\theta$ when using $W_\theta$.
For other directions the modulation is more complex and
only waves propagating near the longitude and latitude directions
are studied here.
The velocity projection also means that waves
traveling in latitude are essentially unobservable
at the equator.
At high latitudes the foreshortening becomes a problem.

\begin{figure}
\plotone{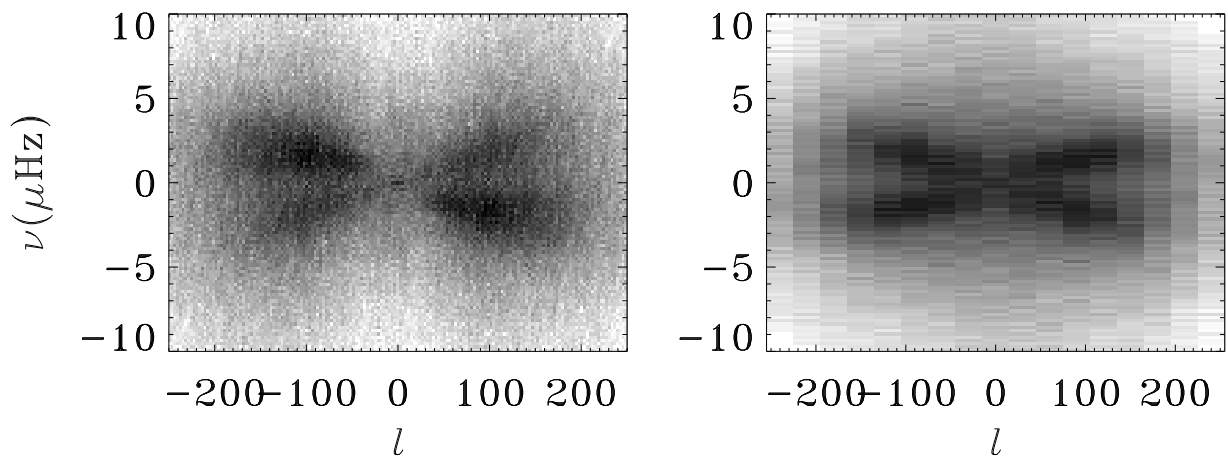}
\caption{
Left: Cut in a 3 dimensional power spectrum using $W_\phi$
at the equator around $k_y=0$.
The grayscale is logarithmic with black 300 times the power of white.
Right: Cut at 40$^\circ$ latitude around $k_x=0$ using $W_\theta$.
The poor resolution is due to the narrow strips in latitude.
The power is symmetric with respect to a simultaneous
change of sign of $l$ and $\nu$.
\label{plotx}}
\end{figure}

Figure \ref{plotx} shows examples of the resulting power spectra.
Note the different power
level in prograde waves (lower right) and retrograde waves (upper
right) in the left panel. In the right panel there is
excess power in the southbound waves (upper right) relative
to the northbound waves (lower right).

The advection of a wave pattern will, at a given azimuth,
lead to two peaks in the power spectra at frequencies
$\omega_\pm = \pm \omega_0(k) + kv$, where $\omega_0(k)$ is the mode
frequency, $k$ is the wavenumber (given here in terms
of the equivalent degree $l=kR_\odot$, where $R_\odot$ is the solar radius)
and $v$ is the component
of the velocity in the direction of propagation.

To quantify the properties of the waves, power spectra at selected
$k$ were extracted and a model consisting
of two Lorentzians and a background
\begin{equation}
M(\omega)=
{{A_-} \over {1+\left ( {{\omega-\omega_-} \over w} \right )^2}}+
{{A_+} \over {1+\left ( {{\omega-\omega_+} \over w} \right )^2}}+
B,
\end{equation}
was fitted to them using a maximum likelihood algorithm.
Here $A_\pm$ are the amplitudes, $w$ is the the (common) width of
the modes and $B$ is a constant background,
The dispersion relation
may be derived as $\omega_0 = (\omega_+ - \omega_-)/2$ and the
flow velocity from $v = (\omega_+ + \omega_-)/2k$.
To account for the projection factor and the
finite observation time, the model
is convolved with an estimate of the appropriate window function.

The power spectra are binned by a factor
of 5 in $k_x$, rebinned to a uniform grid in $k$ and azimuth
and directions within $\pm 15^\circ$ of the desired direction
are averaged before being fitted. This averaging
ensures that the log of the resulting power spectra is close
to normally distributed with a uniform variance, allowing for
a simple maximum likelihood algorithm and stable fits.
The inaccuracy in the parameters caused by the averaging over azimuth
is negligible.

Figure \ref{disp} shows the dispersion relation for the waves
traveling in longitude at the equator.
The solid line shows the fit of a power law over the interval
$50 \le l \le 200$, which gives an exponent of 0.56.
Points below $l\approx 30$ have proven difficult to fit and may
be unreliable.

\begin{figure}
\plotone{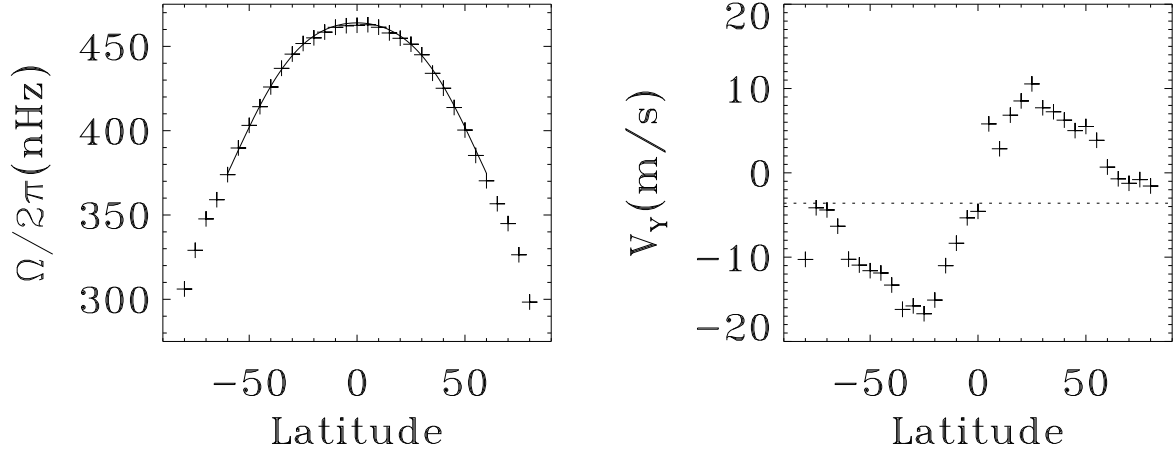}
\caption{Derived flows averaged over $100 \le l \le 200$.
Left: The sidereal rotation rate derived using $W_\phi$ (plusses)
and the rotation rate from \citet{komm93} (solid).
Right: The meridional flow (plusses) (using $W_\theta$)
and an estimate of the effect of the incorrect P-angle (dotted).
\label{flow}}
\end{figure}

Figure \ref{disp} also shows the rotation rate at the equator,
which is almost
independent of $l$, allowing the results to be averaged over $l$,
as done in Figure \ref{flow}, which also shows
the meridional flows.

Finally Figures \ref{power} and \ref{anis} shows the amplitude, linewidth
and anisotropy.

\section{Discussion}

Possibly the main result is to confirm the results of GDS.
Given the differing analysis techniques this represents a significant
confirmation and lays to rest the concerns based on the fact that the
phenomenon had not been observed previously in surface Doppler
shift data.

\begin{figure}
\plotone{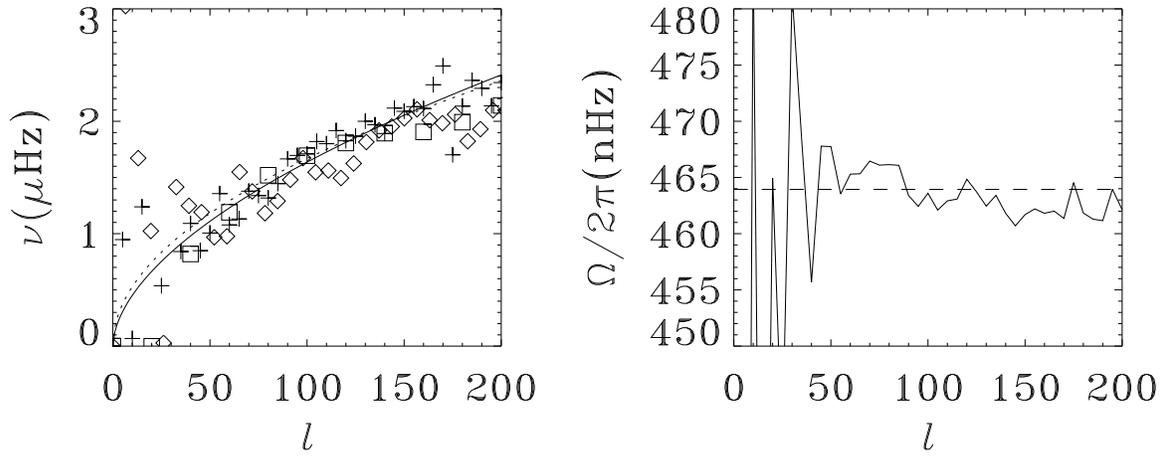}
\caption{
Results for waves traveling in longitude at the equator.
Left: The dispersion relation (plusses),
$1.18(l/50)^{0.5}\mu$Hz (dotted) and
$1.11(l/50)^{0.56}\mu$Hz (solid).
Also shown are the results at $40^\circ$ for waves traveling in longitude
(diamonds) and in latitude (squares). The results at high $l$ are
less reliable due to the foreshortening. Given the poor resolution
in $k_x$ (see Fig. \ref{plotx}) only the results at every 20 in $l$ are
shown for the waves travelign in latitude.
Right: The sidereal rotation rate (solid)
and the equatorial rate from \citet{komm93} (dashed).
\label{disp}}
\end{figure}

While not physically motivated it is interesting that 
the frequencies appear to follow a square root quite well.
As can be seen from Figure \ref{disp} the dispersion relation at
$40^\circ$ is quite similar to that at the equator in both directions.

The small variation of the rotation rate with $l$ is consistent
with a simple depth independent flow. This is in stark contrast to
the numbers derived from the same data in BS,
which varied by 10nHz between $l$=50 and $l$=200,
and which were difficult to explain.
The rotation rate derived here is remarkably close to the magnetic tracer
rate of \citet{komm93}.
The similarity may be related to the fact that the advection of
the small magnetic elements appear to be influenced by the supergranulation.
This similarity is quite unlike the results in BS, in which the
rotation rate,
at least at low $l$, exceeded that of the plasma at any depth.
This lends further credence to the idea that the flows inferred here
represent physical advection rates.

The meridional flow results appear to be reliable to at least
$\pm70^\circ$ latitude with a clear turnover between $20^\circ$ and $30^\circ$
and no sign of a second cell.
Note that the meridional flow is not perfectly
antisymmetric across the equator.
As shown this is largely explained by a P angle error caused by
the misalignment of the MDI instrument on the SOHO spacecraft \citep{toner}
and by the difference between the Carrington elements and the
true rotation axis of the Sun \citep{giles}.
An error in the rotation axis causes part of the rotation velocity to be
misidentified as meridional flow.

\begin{figure}
\plotone{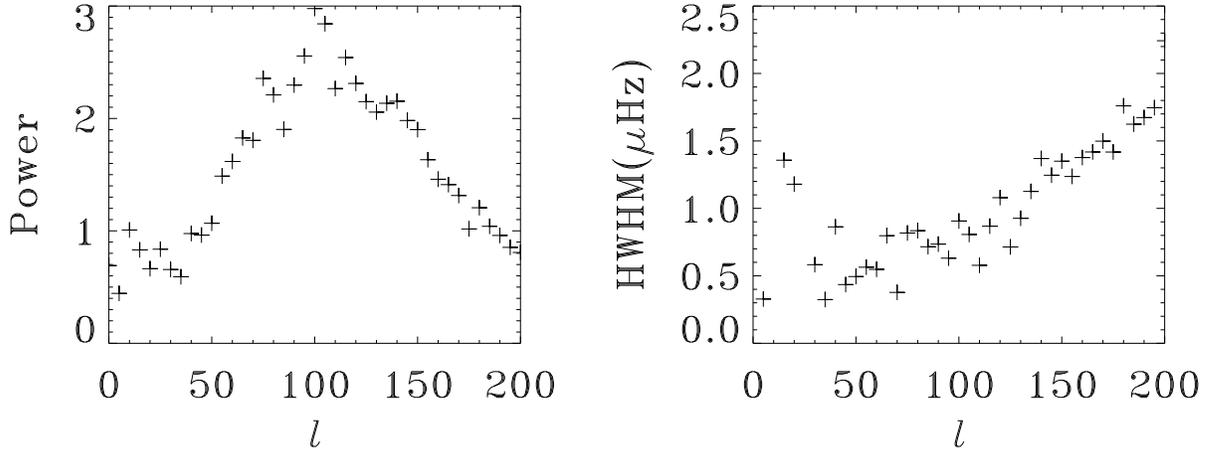}
\caption{
Power $w(A_+ + A_-)/2$ and linewidth $w$ as a function of $l$
for longitudinally traveling waves at the equator.
\label{power}}
\end{figure}

Figure \ref{power} shows that the power peaks around $l=100$, as
previously seen. The linewidth appears
to be an increasing function of $l$, possibly indicating that
smaller scale features damp more quickly. The increase for $l \le 50$
is likely an artifact of the difficulty in fitting low degree features.

\begin{figure}
\plotone{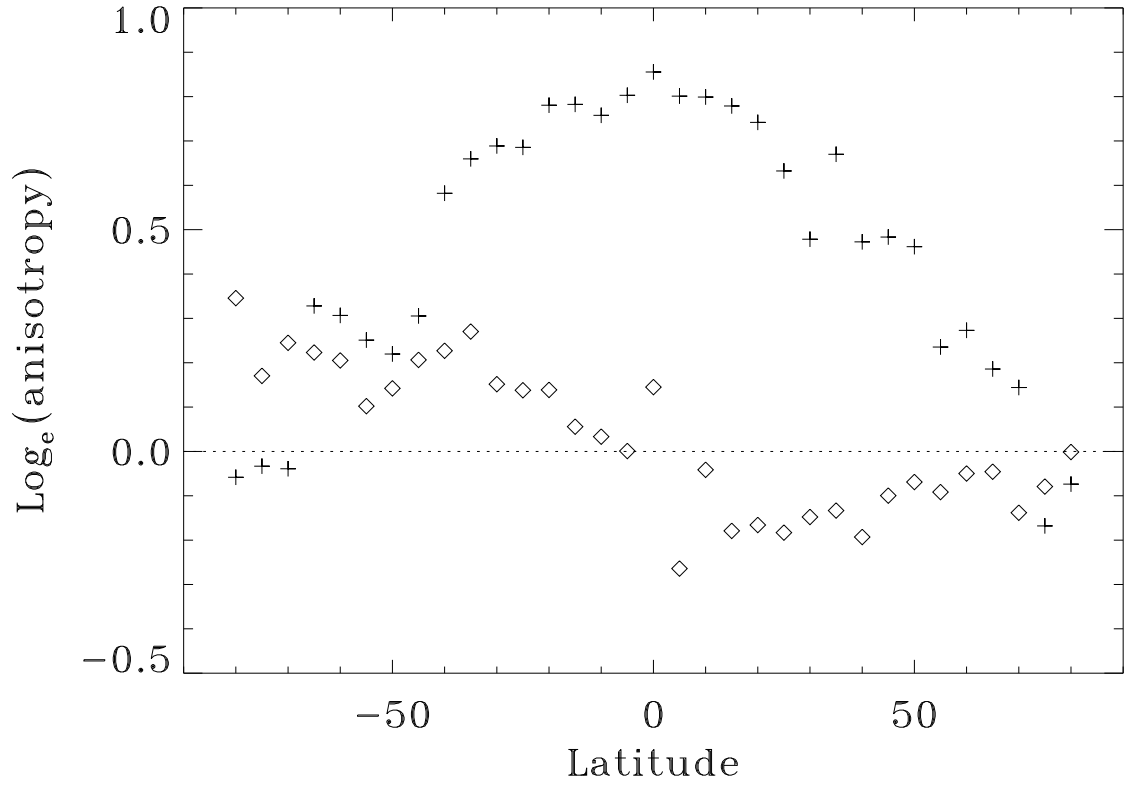}
\caption{
Anisotropy as a function of latitude averaged over $100 \le l \le 200$.
Plusses show the ratio of prograde to retrograde power.
Diamonds show the ratio of Northbound power to Southbound power.
The anisotropy is essentially independent of $l$.
\label{anis}}
\end{figure}

As in GDS Figure \ref{anis} shows that the
power is highly anisotropic at all latitudes and that the anisotropy
is in the prograde direction and increasingly towards the equator at higher
latitudes.

The physical nature of these waves, if that is indeed what they are,
is still unknown. However, with the results presented here, it may
be possible to further constrain the models.

Further progress may be possible by extending these results over time or
by probing the depth structure using time-distance helioseismology.


\acknowledgments

\end{document}